\documentclass[aip,jcp,reprint,amsmath,amssymb,floatfix,citeautoscript]{revtex4-1}
\usepackage{amsmath}
\usepackage{graphicx}
\usepackage{amssymb}
\usepackage{amsthm}
\usepackage{bm}

\begin{document}

\title{Reduced density matrix hybrid approach:
An efficient and accurate method for adiabatic and non-adiabatic quantum dynamics}

\author{Timothy C. Berkelbach}
\email{tcb2112@columbia.edu}
\affiliation{Department of Chemistry, Columbia University, 3000 Broadway, New York, New York 10027, USA}

\author{David R. Reichman}
\affiliation{Department of Chemistry, Columbia University, 3000 Broadway, New York, New York 10027, USA}

\author{Thomas E. Markland}
\affiliation{Department of Chemistry, Stanford University, 333 Campus Drive, Stanford, California 94305, USA} 

\begin{abstract}

We present a new approach to calculate real-time quantum dynamics in complex systems. The formalism is based on
the partitioning of a system's environment into ``core'' and ``reservoir'' modes, with the former to be treated quantum mechanically
and the latter classically.
The presented method only requires the calculation of the system's reduced density matrix
averaged over the quantum core degrees of freedom which is then coupled to a classically evolved reservoir to treat the
remaining modes. We demonstrate our approach by applying it to the spin-boson problem using the noninteracting blip
approximation to treat the system and core, and Ehrenfest dynamics to treat the reservoir. The resulting hybrid methodology is accurate
for both fast and slow baths, since it naturally reduces to its composite methods in their respective regimes of validity.
In addition, our combined method is shown to
yield good results in intermediate regimes where neither approximation alone is accurate and to perform equally well for both strong and weak
system-bath coupling. Our approach therefore provides an accurate and efficient
methodology for calculating quantum dynamics in complex systems.

\end{abstract}

\maketitle

\section{Introduction}

The calculation of real-time quantum dynamics for large molecular systems is a longstanding problem in chemical physics.
Of particular interest is the case of a small subsystem embedded in a surrounding thermal bath, 
which forms the basis for the investigation of condensed phase energy and electron transfer as well as spin and
charge transport in nanoscale devices\cite{wei08,nit06}.
In such situations, one is typically interested in the calculation of the system's {\em reduced} dynamics, averaged over the bath degrees of freedom.
Due to the importance of these problems and the absence of a general solution, a variety of methods have been developed
which vary as to the regimes in which they are accurate and their applicability to large systems.
Many of these approaches rely on either averaging over semi-classical trajectories of bath degrees of freedom or propagating the
system's reduced density matrix (RDM) directly without explicit treatment of the bath, for example by master equations or path integral techniques. 

Trajectory based methods include the mean-field Ehrenfest\cite{tul98,sto95},
surface-hopping\cite{pec69,tul98,tul90}, semi-classical initial value
representation\cite{mil01,mil09}, linearization\cite{shi03,pou03,bon05,dun08}, and classical mapping\cite{mey79,sto97,tho99}
approaches. These offer the appeal of a transparent link to the classical limit and
simple application to complex condensed phase systems and to slow baths.
However, these methods typically fail in regimes with strong system-bath coupling
or high-frequency baths where quantum effects are known to be important.

Techniques based on direct propagation of the reduced density matrix include the Markovian and non-Markovian
Bloch-Redfield equations\cite{blo57,red65,bre02} and the noninteracting blip approximation\cite{leg87,wei08}
which serves as a time-dependent generalization of F\"{o}rster theory\cite{for53}. To make such approaches
manageable, perturbation theory in the system-bath coupling (Bloch-Redfield) and in the non-adiabatic
coupling (NIBA, F\"{o}rster) are typically carried out only to second-order and hence performance degrades
when these terms become large. In addition, path integral approaches, such as quantum Monte Carlo \cite{mak89,mak91,egg94}
and the quasi-adiabatic propagator path integral (QUAPI)\cite{mak92} have also been developed to propagate
the RDM. While these approaches can be made numerically exact, they are frequently difficult to converge in
practice when the system-bath coupling is strong or the bath is slow.  Furthermore, many variants scale poorly
with the number of system states, drastically limiting the size of treatable systems.

Hence when the bath dynamics are ``fast'' compared to those of the system (the so-called {\em non-adiabatic} regime), both perturbative
quantum master equations and numerically exact path integral approaches are typically accurate since the correlation time
of the bath is short, rendering non-Markovian memory effects less significant. When the bath dynamics are ``slow'' compared
to those of the system (the {\em adiabatic} regime), these methods fail but it becomes reasonable to treat the bath degrees
of freedom classically and hence trajectory-based schemes generally yield accurate results. Given this complementarity, it would
clearly be desirable to develop a scheme which can accurately treat both regimes by naturally tuning between the above approaches
in a consistent manner. In addition, such an approach might effectively treat intermediate regimes where
no obvious time-scale separation exists and hence neither RDM nor trajectory-based approaches are accurate.

The ``dynamical hybrid'' scheme of Wang, Thoss and Miller \cite{wan01,tho01} offers such an approach in which bath degrees of
freedom are partitioned into ``core'' modes to be treated with quantum mechanics and ``reservoir'' modes to be treated
with classical mechanics.
In this scheme, the wavefunction-based multi-configurational time-dependent Hartree
(MCTDH)\cite{mey90,man92,bec00} method is used for the system and core while a mean-field Ehrenfest treatment is applied to
the reservoir.
When applied to the spin-boson problem, this dynamical hybrid scheme frequently converges to the exact result
with only a small percentage of modes included in the core.
However, to achieve convergence in the presence of fast baths, up to 80\% of the bath modes need to be included in the core.
Due to the high cost of treating modes using MCTDH, this severely reduces the efficiency of the dynamical hybrid approach and hence
limits the ease with which it can be applied to larger systems.

In this work, we present a methodology which avoids the use of wavefunctions, replacing them with a single
reduced density matrix averaged over the quantum core modes.  The reservoir modes exert a classical
time-dependent driving force on this system RDM, which in turn drives the classical modes via a back-reaction force.
This reduced density matrix hybrid formalism (RDM-Hybrid)
allows one to use an appropriately chosen partitioning of the bath into core and reservoir modes such that
the system RDM can be propagated either by an exact reduced dynamics algorithm {\em with a decreased cost}
or by an inexpensive approximate master equation {\em with improved accuracy}.
In other words, by using physical intuition, one can choose the core-reservoir partitioning such that different methods 
are applied only to the set of bath modes for which they are expected to work.

We demonstrate our approach by applying it to the
spin-boson model and treating the system and core using the noninteracting blip approximation (NIBA)\cite{leg87,wei08}
This approach yields a methodology which is numerically cheap while obtaining impressive quantitative accuracy over all
regimes investigated as well as offering excellent scaling with the number of system states.  

The outline of the paper is as follows.  Section~\ref{ssec:ehren} briefly reviews the Ehrenfest methodology.
This background is then used in Sec.~\ref{ssec:hybrid}, to derive the RDM-Hybrid approach.
Section~\ref{sec:spinboson} then describes how this scheme can be applied to the spin-boson model.
The results of the application are presented in Sec.~\ref{sec:results}, and Sec.~\ref{sec:conc} concludes.

\section{Theory}\label{sec:theory}

\subsection{Ehrenfest dynamics}\label{ssec:ehren}

While there are many routes to the mean-field Ehrenfest equations of motion and subsequently the
RDM-Hybrid method to be presented below, we proceed via the quantum-classical Liouville equation, which provides
a particularly simple derivation.

We begin by considering the generic coupled system-bath Hamiltonian,
\begin{equation}
H = H_s(s,p_s) + H_b(\mathbf{Q},\mathbf{P}) + V_{sb}(s,\mathbf{Q})
\end{equation}
where the free bath Hamiltonian is
\begin{equation}
H_b(\mathbf{Q},\mathbf{P}) = \sum_k \left[ \frac{P_k^2}{2} + V(Q_k) \right]
\end{equation}
and the coupling is assumed to be of the form
\begin{equation}
V_{sb} = \sum_i F_i(s) G_i(\mathbf{Q}).
\end{equation}

The dynamics of the total density matrix, $\rho(t)$, is given by the Liouville equation,
\begin{equation}
\frac{\partial \rho(t)}{\partial t} = -i \big[ H, \rho(t) \big],
\end{equation}
where henceforth we set $\hbar = 1$.
By employing the partial Wigner transform of the density matrix with respect to the $f$ environmental
degrees of freedom,
\begin{align}
\rho^W(\mathbf{Q},\mathbf{P},t) & = \frac{1}{(2\pi)^f} \int d^f \mathbf{y} e^{-i \mathbf{P}\cdot \mathbf{y}} \nonumber\\
		&\hspace{5em} \times \langle \mathbf{Q} + \frac{\mathbf{y}}{2} | \hat{\rho} | \mathbf{Q} - \frac{\mathbf{y}}{2}\rangle,
\end{align}
and taking the classical limit of the corresponding partial Wigner transformed Liouville equation
(for more details, see e.g. Ref.~\onlinecite{kap06}), one obtains the so-called
quantum-classical Liouville equation
\begin{align}
&\frac{\partial \rho^W(\mathbf{Q},\mathbf{P},t)}{\partial t} = -i \big[ H^W,\rho^W(t) \big] \nonumber\\
&\hspace{4em} + \frac{1}{2} \left( \big\{ H^W,\rho^W(t) \big\} + \big\{ \rho^W(t), H^W \big\} \right),
\end{align}
where $[\cdot,\cdot]$ denotes a commutator and $\{\cdot,\cdot\}$ is the Poisson bracket.

Now, one makes the approximation
that the system and bath reduced density matrices (RDMs) decouple at all times,
\begin{equation}
\rho^W(\mathbf{Q},\mathbf{P},t) = \rho_s(t) \rho_b(\mathbf{Q},\mathbf{P},t),
\end{equation}
with the system RDM $\rho_s(t) = \int d^f \mathbf{Q} \int d^f\mathbf{P} \rho^W (\mathbf{Q},\mathbf{P},t)$
and bath RDM $\rho_b(\mathbf{Q},\mathbf{P},t) = {\rm Tr}_{\rm sys} \big[ \rho^W(t) \big]$.
Inserting this product form into the quantum-classical Liouville equation and noting that the
bath reduced density matrix is simply the classical phase space distribution,
\begin{equation}
\rho_b(\mathbf{Q},\mathbf{P},t) = \delta(\mathbf{Q}-\mathbf{Q}(t))\delta(\mathbf{P}-\mathbf{P}(t))
\end{equation}
one obtains the mean-field Ehrenfest equations
of motion\cite{gru09} for the system RDM,
\begin{align}
\frac{\partial \rho_s(t)}{\partial t} &= -i \big[ H_s + V_{sb}\big(s,\mathbf{Q}(t)\big), \rho_s(t) \big] \nonumber\\
	& = -i \big[ H_s + \sum_i F_i(s) G_i\big(\mathbf{Q}(t)\big), \rho_s(t) \big] \nonumber\\
	& \equiv -i \big[ \tilde{H}_{s}(t), \rho_s(t) \big],
\end{align}
and the classical bath degrees of freedom,
\begin{align}
\frac{d Q_{k}}{dt} &= P_{k} \\
\frac{d P_{k}}{dt} &= -\frac{\partial}{\partial Q_k} \Big[ V(Q_k) + {\rm Tr}_{s} \big\{ V_{sb}\big(s,\mathbf{Q}(t)\big) \big\} \Big] \nonumber\\
	& = -\frac{\partial V(Q_k)}{\partial Q_k}
	- \sum_i \frac{\partial G_i\big(\mathbf{Q}(t)\big)}{\partial Q_k} {\rm Tr}_{s} \big\{ F_i \rho_s(t) \big\}.
\end{align}

The Ehrenfest approximation to an observable of the reduced system is given by
\begin{align}
& \langle O_s(t) \rangle \equiv {\rm Tr}_s{\rm Tr}_b \left\{ O_s \rho(t) \right\}
	\approx {\rm Tr}_b \Big\{ {\rm Tr}_s \left\{ O_s \rho_s(t)\right\} \rho_b(t) \Big\} \nonumber\\
&\hspace{2em} \approx \int d^f \mathbf{Q} \int d^f \mathbf{P} {\rm Tr}_s \left\{ O_s \rho_s(t)\right\} \rho_b (\mathbf{Q},\mathbf{P},t).
\end{align}
In accordance with the classical treatment of the bath, this latter integral is evaluated by molecular dynamics
obeying the equations of motion given above.  To generate the canonical ensemble average consistent with the initial density matrix
\begin{equation}
\rho(0) = \rho_s(0) \rho_b(0) = \rho_s(0) \exp(-\beta H_b) / Z_b,
\end{equation}
one averages over classical trajectories with bath initial conditions sampled from either the Wigner-transformed
bath density operator,
\begin{align}
\rho^W_b(\mathbf{Q}_0,\mathbf{P}_0) & = \frac{1}{(2\pi)^f} \int d^f \mathbf{y} e^{-i \mathbf{P}_0\cdot \mathbf{y}} \nonumber\\
		&\hspace{5em} \times \langle \mathbf{Q}_0 + \frac{\mathbf{y}}{2} | \hat{\rho}_b | \mathbf{Q}_0 - \frac{\mathbf{y}}{2}\rangle,
\end{align}
or more simply from the classical Boltzmann distribution,
\begin{equation}
\rho^B_b(\mathbf{Q}_0,\mathbf{P}_0) = e^{-\beta H_b(\mathbf{Q}_0,\mathbf{P}_0)} / Z_b.
\end{equation}
Although the Wigner distribution gives initial conditions consistent with the exact quantum distribution,
this property is not conserved by the classical dynamics.

\subsection{Reduced density matrix hybrid dynamics}\label{ssec:hybrid}

In contrast to the approach taken above, in which the full Hamiltonian is split into a system and bath with coupling between the two,
we instead divide the bath degrees of freedom into two sets:
the reservoir modes ($\{Q_k,P_k : k=1,\dots,f^\prime\}$) to be treated classically and the core modes
($\{q_k,p_k : k = f^\prime+1,\dots,f\}$) to be treated quantum mechanically. Hence, we partition the Hamiltonian as
\begin{equation}\label{eq:partition}
H = H_{sc}(s,p_s,\mathbf{q},\mathbf{p}) + H_r(\mathbf{Q},\mathbf{P}) + V_{sr}(s,\mathbf{Q})
\end{equation}
where the system-core Hamiltonian is
\begin{equation}
H_{sc}(s,p_s,\mathbf{q},\mathbf{p}) = H_s(s,p_s) + H_c(\mathbf{q},\mathbf{p}) + V_{sc}(s,\mathbf{q})
\end{equation}
and the system-core and system-reservoir couplings are defined as
\begin{align}
V_{sc}(s,\mathbf{q}) &= \sum_i F_i(s)G_i(\mathbf{q}), \\
V_{sr}(s,\mathbf{Q}) &= \sum_i F_i(s)G_i(\mathbf{Q}).
\end{align}

Since nothing in our above discussion of Ehrenfest dynamics utilized the specific properties of the system,
we could just as well re-label the {\em combined system and core} as the {\em system} of the previous section.
Likewise, we associate the {\em reservoir} with the {\em bath} from before.
Thus we assume the total (Wigner-transformed) density matrix factorizes at all times into a system-core RDM and reservoir RDM,
\begin{equation}
\rho^W(t) = \rho_{sc}(t)\rho_b(\mathbf{Q},\mathbf{P},t).
\end{equation}
Taking the classical limit we arrive at the Ehrenfest-like
RDM-Hybrid equations of motion for the system-core density matrix,
\begin{align}
\frac{\partial \rho_{sc}(t)}{\partial t} &= -i \big[ H_{sc} + V_{sr}\big(s,\mathbf{Q}(t)\big), \rho_{sc}(t) \big] \nonumber\\
	& = -i \big[ H_s + \sum_i F_i(s) G_i\big(\mathbf{Q}(t)\big), \rho_{sc}(t) \big] \nonumber\\
	& \equiv -i \big[ \tilde{H}_{s}(t), \rho_{sc}(t) \big], \label{eq:rhosc}
\end{align}
and the classical bath degrees of freedom,
\begin{align}
\frac{d Q_{k}}{dt} &= P_{k} \label{eq:batheom1}\\
\frac{d P_{k}}{dt} &= -\frac{\partial}{\partial Q_k} \Big[ V(Q_k) + {\rm Tr}_s {\rm Tr}_c \big\{ V_{sb}\big(s,\mathbf{Q}(t)\big) \big\} \Big] \nonumber\\
	& = -\frac{\partial V(Q_k)}{\partial Q_k}
	- \sum_i \frac{\partial G_i\big(\mathbf{Q}(t)\big)}{\partial Q_k} {\rm Tr}_s {\rm Tr}_c \big\{ F_i \rho_{sc}(t) \big\}. \label{eq:batheom2a}
\end{align}
Although the exact calculation of the system-core density matrix dynamics given by Eq.~(\ref{eq:rhosc})
is extremely demanding, this was essentially the approach
taken by Wang, Thoss, and Miller (WTM)\cite{wan01,tho01} who averaged high-dimensional wavefunction trajectories with initial conditions
of the core wavefunction sampled from the exact quantum mechanical core density matrix and classical reservoir degrees of freedom sampled
from the quantum-classical Wigner distribution.

The important point we seek to make in this work is that $F_i$ is a pure system operator, i.e. the core and reservoir are not directly coupled,
such that the classical reservoir equation of motion above may be written
\begin{align}
\frac{d P_{k}}{dt} = -\frac{\partial V(Q_k)}{\partial Q_k}
	- \sum_i \frac{\partial G_i\big(\mathbf{Q}(t)\big)}{\partial Q_k} {\rm Tr}_s \big\{ F_i \rho_s(t) \big\}, \label{eq:batheom2b}
\end{align}
where we have introduced the system RDM, $\rho_s(t) = {\rm Tr}_c \big\{ \rho_{sc}(t) \big\}$, averaged over the quantum core degrees
of freedom.  Furthermore, the calculation of any
dynamical system variable is given by an expression analogous to the pure Ehrenfest approximation above,
\begin{align}
& \langle O_s(t) \rangle \equiv {\rm Tr}_s{\rm Tr}_c{\rm Tr}_r \left\{ O_s \rho(t) \right\}
	\approx {\rm Tr}_r \Big\{ {\rm Tr}_s \left\{ O_s \rho_s(t)\right\} \rho_r(t) \Big\} \nonumber\\
&\hspace{1em} \approx \int d^{f^\prime} \mathbf{Q} \int d^{f^\prime}
	\mathbf{P} {\rm Tr}_s \left\{ O_s \rho_s(t)\right\} \rho_r (\mathbf{Q},\mathbf{P},t),
\end{align}
which again only requires the system RDM (and {\em not} the much higher-dimensional system-core RDM), with classical reservoir trajectories
to be sampled from an appropriate distribution.

We have thus arrived at a self-consistent dynamical scheme which treats the core bath modes quantum mechanically
yet only requires the propagation of a typical system RDM and classical reservoir degrees of freedom.
This RDM-Hybrid formulation presented above constitutes the major theoretical result of this work. 
We point out that a similar approach appears to have been
investigated by Golosov et al.\cite{gol00_fries}, though it was restricted to their
previously developed memory-equation algorithm\cite{gol99} and only minimally pursued.

When the system RDM averaged over the quantum core modes, $\rho_s(t)$, is treated fully quantum mechanically,
our methodology yields the exact result when all bath modes are included in the core.  However,
in practice one expects to achieve numerical convergence before this limit is reached.  Of course, the exact quantum
dynamical treatment of the system RDM is in general no less numerically challenging than
the original problem (a system coupled to a quantum bath).
However, the advantage of this approach is that one can tailor the partitioning in Eq.~(\ref{eq:partition})
either to alleviate the computational expenses of
a numerically exact reduced quantum dynamics method (such as the QUAPI-based iterative tensor propagation scheme of
Makri et al.\cite{mak92,mak94,mak95_1,mak95_2})
or to improve the accuracy of approximated quantum dynamics (such as a perturbative quantum master equation) for the system RDM.
In this sense, the above presented formalism outlines a {\em framework} for the efficient partitioning and calculation of real-time
reduced quantum dynamics where one can employ any exact or approximate dynamical scheme for the system RDM depending on the
complexity of the problem. As discussed in the following section, here we adopt an approximate treatment of the system RDM
using a perturbative master equation to yield a very cheap hybrid methodology with impressive quantitative accuracy.
 
\section{Application to the spin-boson problem}\label{sec:spinboson}
While the above formalism is indeed quite general here we demonstrate its effectiveness at treating the spin-boson Hamiltonian,
\begin{equation}\label{eq:spinboson}
H = \varepsilon \sigma_z + \Delta \sigma_x + \sigma_z \sum_k c_k Q_k + \sum_k \left[ P_k^2 + \omega_k^2 Q_k^2 \right]/2, 
\end{equation}
which describes a two-level system with energy bias $2\varepsilon$ and tunneling matrix element $\Delta$, linearly coupled
to the coordinates of a bath of harmonic oscillators.  The spin-boson Hamiltonian has been used as a model for
a variety of physically distinct quantum relaxation and transport processes. 

A principal observable of interest is the population variable,
\begin{equation}\label{eq:population}
P(t) = {\rm Tr}_{\rm sys} {\rm Tr}_{\rm bath}\left[ \sigma_z(t) \rho(0) \right] \equiv \langle \sigma_z(t) \rangle,
\end{equation}
which measures the difference in population of sites 1 and 2.  Furthermore, we will henceforth work in reduced units 
with respect to the tunneling matrix element, $\Delta$, in addition setting $\hbar = k_B = 1$.
To begin our application, we first discuss the separation of modes into the core and the reservoir.

\subsection{Separation of modes}

The harmonic bath in the spin boson problem is fully specified by its spectral density
\begin{equation}
J(\omega) = \frac{\pi}{2} \sum_k \frac{c_k^2}{\omega_k} \delta(\omega-\omega_k).
\end{equation}
When separating the bath modes into a core and reservoir, we will require that
\begin{equation}
J_{\rm core}(\omega) + J_{\rm res}(\omega) = J(\omega).
\end{equation}
Intending to treat high frequency modes in the quantum mechanical core and low frequency modes in the classical reservoir,
we use a switching function, $S(\omega,\omega^*)$,
which switches from 1 to 0 as $\omega$ goes from 0 to $\omega^*$.  This allows the core and reservoir spectral densities to be defined as,
\begin{align}
J_{\rm core}(\omega) &= J(\omega)\left[1 - S(\omega,\omega^*)\right], \\
J_{\rm res}(\omega) &= J(\omega)S(\omega,\omega^*).
\end{align}
An infinite number of switching functions satisfy such a criteria and the best form will depend on the Hamiltonian adopted,
the methodology used to treat the core, and the required balance been accuracy and efficiency.
The simplest approach would be to use a step function,
\begin{equation}\label{eq:step}
S(\omega,\omega^*) = \begin{cases}
1 & \omega < \omega^* \\
0 & \omega > \omega^*.
\end{cases}
\end{equation}
However in many cases, such a sharp switch will leave a core which is expensive to treat using quantum mechanics.
The reason for this difficulty can be seen by considering the bath correlation function, given by
\begin{align}
&\alpha(t) = \sum_k c_k^2 \langle Q_k(t) Q_k(0) \rangle \nonumber\\
& =\frac{1}{\pi} \int_0^\infty d\omega J(\omega) \left[ \coth(\beta\omega/2) \cos(\omega t) - i \sin(\omega t)\right],
\end{align}
whose correlation time determines the range of memory (non-Markovian) effects.
It is straightforward to see that using a step function switching like that in Eq.~(\ref{eq:step}) will result in a
long-time oscillatory tail in the bath correlation function, $\alpha(t)$, and thus extensive non-Markovian effects which are difficult
to treat quantum mechanically. It is therefore advantageous to separate the spectral density using a smooth switching function which can yield
a short bath correlation time for the core modes and allow for efficient quantum mechanical treatment. 

\begin{figure}[t]
\centering
\includegraphics[scale=0.4]{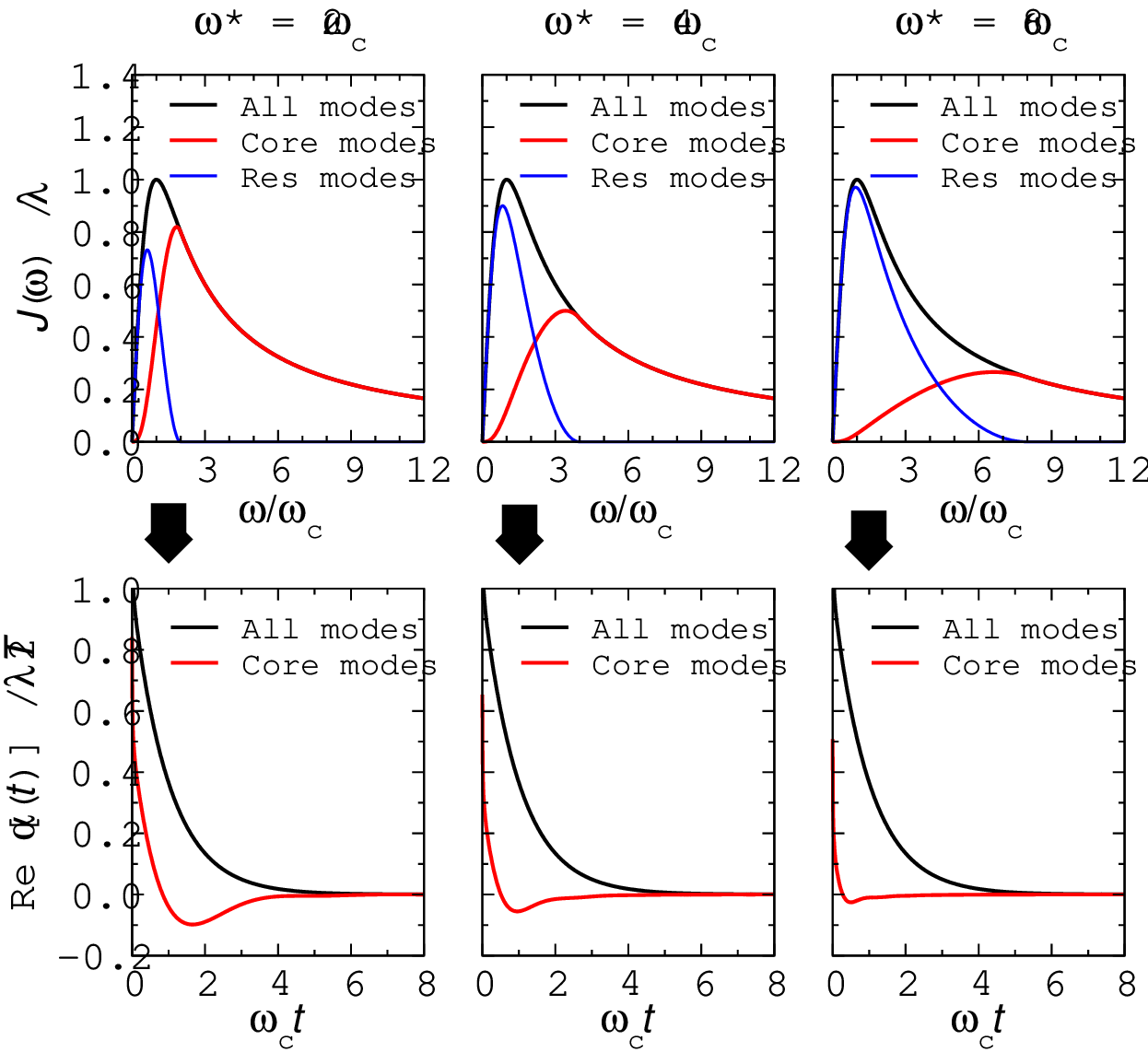}
\caption{The effect of splitting the spectral density (top) on the resulting bath correlation functions with $T/\omega_c = 4$ (bottom).}
\label{fig:split}
\end{figure}

For our purposes a switching function of the form,
\begin{equation}\label{eq:smoothswitch}
S(\omega,\omega^*) = \begin{cases}
\left[1 - \left(\omega/\omega^*\right)^2\right]^2 & \omega < \omega^* \\
0 & \omega > \omega^*
\end{cases}
\end{equation}
was chosen. In Fig.~\ref{fig:split}, we show the bath correlation function when this function is applied to the Debye spectral density,
\begin{equation}\label{eq:debye}
J(\omega) = 2 \lambda \omega_c \frac{ \omega }{\omega^2 + \omega_c^2},
\end{equation}
where $\lambda = \pi^{-1} \int_0^\infty d\omega J(\omega)/\omega$ is the reorganization energy. As can be seen,
this smooth switching does not introduce any long-time tails and allows the bath correlation time for the core
modes to be decreased as $\omega^*$ is increased and more of the slow modes, which give rise to temporally non-local
effects, are moved into the reservoir. In addition, removal of slow modes from the core encourages the use
of approximate master equations, such as the noninteracting blip approximation discussed in the next section,
which complement the Ehrenfest treatment of the slow modes. 

\subsection{Noninteracting blip approximation for the system and core}

Although the RDM-Hybrid framework allows for the application of many approaches to treat the system and core modes,
in this work we employ the noninteracting blip approximation (NIBA)\cite{leg87,wei08}.  To see why such a choice is computationally simple,
note that because the fluctuating classical degrees of freedom provide an effective time-dependent bias,
the system Hamiltonian no longer commutes with itself at different times, and thus its propagator is a cumbersome
time-ordered exponential (Dyson series).  However, since NIBA only treats the diagonal part of this Hamiltonian exactly,
its propagator is trivially calculated. 

\begin{figure}[t]
\centering
\includegraphics[scale=0.4]{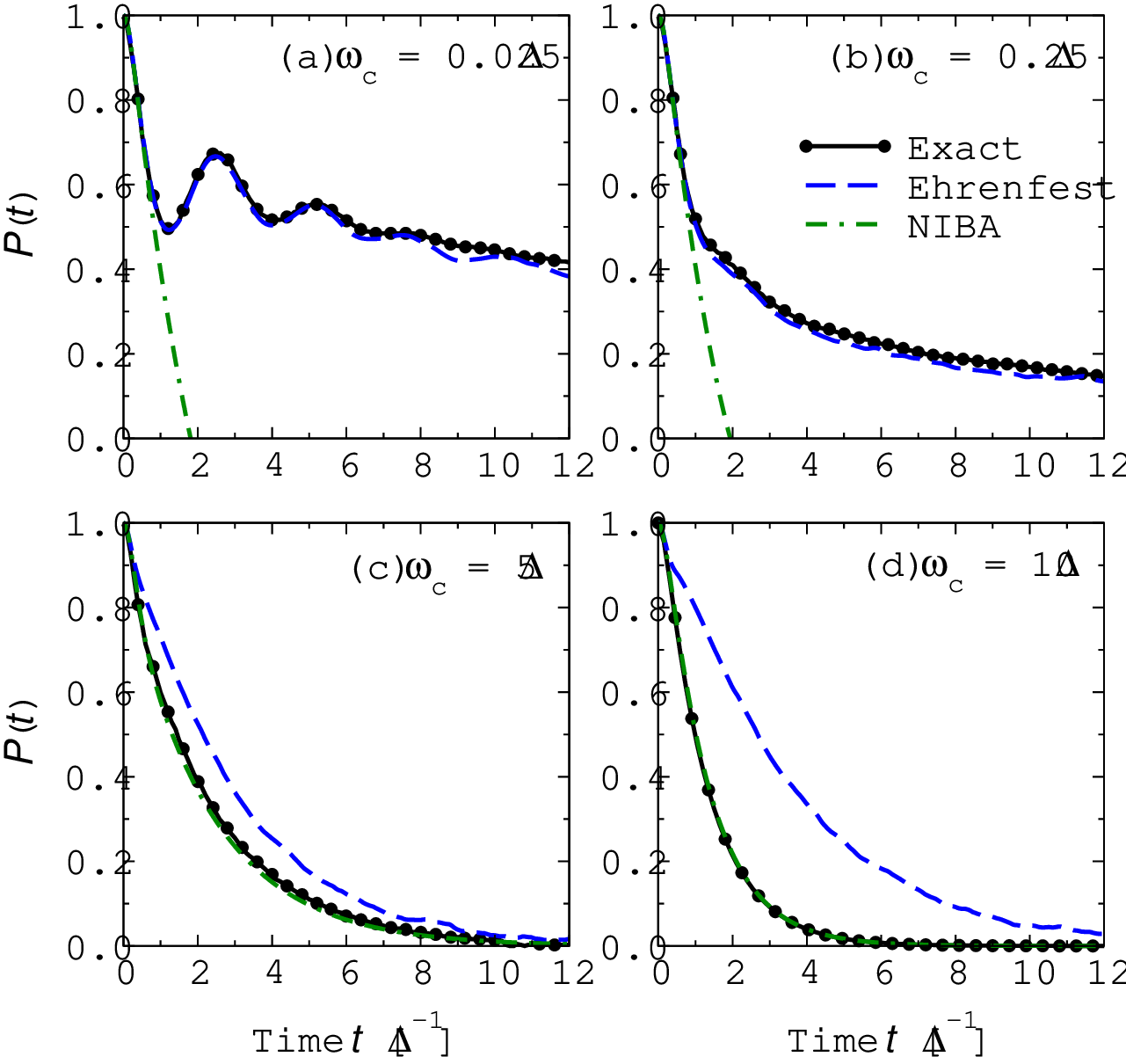}
\caption{Population dynamics, $P(t) = \langle \sigma_z(t) \rangle$, for the spin-boson model with a Debye spectral density.
The Hamiltonian parameters are
$\varepsilon = 0$, $\lambda = 2.5\Delta$, and $T = 2\Delta$.  Ehrenfest and NIBA are compared to the numerically exact
results of Thoss, Wang, and Miller\cite{tho01} in (a)-(c) and to our own converged QUAPI calculation in (d).}
\label{fig:crossover}
\end{figure}

Another important reason for our choice is the complementarity between NIBA and Ehrenfest. While Ehrenfest dynamics
work best in the adiabatic regime, $\omega_c/\Delta \ll 1$, NIBA is perturbative in $\Delta/\omega_c$,
thus working best in the non-adiabatic regime, $\omega_c/\Delta \gg 1$.
To exemplify this complementarity, in
Fig.~\ref{fig:crossover} we show a coherent to incoherent crossover, as a function of $\omega_c/\Delta$ in the high-temperature,
strong-coupling regime. As can be seen in Fig.~\ref{fig:crossover}, the Ehrenfest method yields quantitatively exact population
dynamics for an adiabatic choice of parameters, $\omega_c/\Delta \ll 1$ [panels
(a) and (b)], whereas NIBA gives the exact result upon crossing over into the non-adiabatic regime,
$\omega_c/\Delta \gg 1$ [panels (c) and (d)].
With this in mind we are strongly encouraged to consider a NIBA-Ehrenfest
hybridization, the details of which are described next.

\subsection{Simulation details}

The total bath was discretized by the relation,
\begin{equation}
c_k^2 = \frac{2}{\pi} \omega_k \frac{J(\omega_k)}{\rho(\omega_k)},
\end{equation}
where $\rho(\omega)$ is a density of frequencies chosen to reproduce the reorganization energy for any number of modes, $f$.  For
an Ohmic spectral density $J(\omega) = \lambda \omega/\omega_c F(\omega/\omega_c)$ with cutoff function $F(\omega/\omega_c)$,
we thus require
\begin{align}
\lambda &= \frac{1}{\pi} \int_0^\infty d\omega \frac{J(\omega)}{\omega} = \frac{1}{2} \sum_{k=1}^{f} \frac{c_k^2}{\omega_k^2} \nonumber\\
	&= \frac{\lambda}{\pi \omega_c} \sum_{k=1}^{f} \frac{F(\omega_k/\omega_c)}{\rho(\omega_k)},
\end{align}
which is clearly satisfied by the choice
\begin{equation}
\rho(\omega) = \frac{f}{\pi\omega_c} F(\omega/\omega_c).
\end{equation}
For the Debye spectral density considered here, the cutoff function is given by
$F(\omega/\omega_c) = 2 / [ 1 + \left(\omega/\omega_c\right)^2 ]$.

Having obtained a discrete distribution of frequencies, we then take into account the switching function and {\em recalculate} the
reservoir couplings as
\begin{equation}
c_k^2 = \frac{2}{\pi} \omega_k \frac{J_{\rm res}(\omega_k)}{\rho(\omega_k)},\hspace{2em} k = 1, \dots, f^\prime \leq f
\end{equation}
in the process removing all bath modes with $ \omega_k > \omega^*$ from the reservoir.
In a complimentary fashion, the core spectral density, $J_{\rm core}(\omega)$, is used in the NIBA equations.

In all results presented, $f=300$ discrete bath modes were found to be sufficient for convergence and averages were
performed over $10^3-10^4$ reservoir initial conditions.
For pure Ehrenfest results, we sample from the Wigner distribution, which for the spin-boson Hamiltonian's harmonic bath
takes the form,
\begin{align}
&\rho^W_b(\mathbf{Q},\mathbf{P}) = \prod_{k=1}^{f} \frac{\tanh(\beta \omega_k/2)}{\pi} \nonumber\\
&\hspace{3em} \times \exp \left[ - \frac{2 \tanh(\beta\omega_k/2)}{\omega_k}
			\left(\frac{P_k^2}{2}+\frac{\omega_k^2 Q_k^2}{2} \right)\right].
\end{align}

As alluded to above, when high frequency baths are present, Wigner sampling can improve the
accuracy of Ehrenfest dynamics since the initial conditions are consistent with the exact quantum mechanical distribution.
However, since the zero-point energy inserted by Wigner sampling is not conserved by the classical dynamics, long-time
predictions can be unreliable (zero-point energy leakage)\cite{sto99,mul99}. Hence, in regimes with strong coupling or high bath frequencies,
sampling from either the classical Boltzmann or quantum Wigner distributions
can produce inaccurate results. In contrast, for our RDM-Hybrid approach we find that the results are
largely insensitive to the choice of reservoir initial conditions, since most quantum mechanical modes are included in the
core leading to only low frequencies being present in the reservoir, for which classical and Wigner sampling give identical results.
This effect is shown later, in Fig.~\ref{fig:nonadiabatic}, where pure Ehrenfest dynamics using classical sampling is
seen to underestimate the decay of the system population variable while using Wigner sampling overestimates the decay.
The RDM-Hybrid approach, however, produces graphically identical results using either sampling scheme.

The time evolution of the coupled classical dynamics and NIBA equations were solved with a second-order Runge-Kutta scheme with a timestep
of $0.01\Delta^{-1}$.  
For the spin-boson Hamiltonian considered here, the reservoir equations of motion, Eqs.~(\ref{eq:batheom1})-(\ref{eq:batheom2b}), take the
form 
\begin{align}
\frac{d Q_{k}}{dt} &= P_{k} \\
\frac{d P_{k}}{dt} &= -\omega_k^2 Q_k
	- c_k \tilde{P}(t),
\end{align}
with the system population variable $\tilde{P}(t) \equiv {\rm Tr}_{s} \left\{ \sigma_z \rho_s(t) \right \} $.
Taking $\tilde{P}(t)$ to be constant over a half-timestep as called for by the Runge-Kutta scheme used here, one has
\begin{align}
Q_k(t+\Delta t/2) &= \left[ Q_k(t) + \frac{c_k}{\omega_k^2} \tilde{P}(t) \right] \cos (\omega_k \Delta t/2) \nonumber\\
		& \hspace{1em} + \frac{P_k(t)}{\omega_k} \sin ( \omega_k \Delta t/2) - \frac{c_k}{\omega_k^2}\tilde{P}(t),\\
P_k(t+\Delta t/2) &= -\left[ Q_k(t) + \frac{c_k}{\omega_k^2} \tilde{P}(t) \right] \omega_k \sin (\omega_k \Delta t/2) \nonumber\\
		& \hspace{1em} + \frac{P_k(t)}{\omega_k} \cos ( \omega_k \Delta t/2).
\end{align}

For a two-level system coupled to a common bath, the NIBA master equation for the population difference
with a time-dependent bias is given simply as\cite{goy95}
\begin{equation}\label{eq:niba}
\frac{d\tilde{P}(t)}{dt} = -\int_0^t d\tau \left[ K_+(t,\tau)\tilde{P}(\tau) + K_-(t,\tau)\right],
\end{equation}
with the kernels
\begin{align}
K_+(t,\tau) & = 4\Delta^2 \exp\left[-Q_2(t-\tau)\right]\cos\left[Q_1(t-\tau)\right] \nonumber\\
		&\hspace{1em}\cos\left\{ \zeta(t,\tau) + (1+\delta)\left[Q_1(t)-Q_1(\tau)\right] \right\}, \label{eq:kernel1}\\
K_-(t,\tau) & = 4\Delta^2 \exp\left[-Q_2(t-\tau)\right]\sin\left[Q_1(t-\tau)\right] \nonumber\\
		&\hspace{1em}\sin\left\{ \zeta(t,\tau) + (1+\delta)\left[Q_1(t)-Q_1(\tau)\right] \right\}. \label{eq:kernel2}
\end{align}

The parameter $\delta$ originates from the bath initial condition, with $\delta = 0$ if the system and bath are initially uncoupled
and $\delta = -1$ if the bath is initially solvating the donor.
The twice-integrated bath correlation function, $Q(t) = Q_2(t) + iQ_1(t)$, is given by
\begin{align}
Q(t) = \frac{4}{\pi} \int_0^\infty d\omega \frac{J_{\rm core}(\omega)}{\omega^2}
		& \left\{ \coth(\beta\omega/2)\left[1-\cos(\omega t)\right]\right. \nonumber\\
	&\left.+\ i\sin(\omega t) \right\}.
\end{align}

The effect of a general time-dependent bias $2\varepsilon(t)$ is captured in the function
\begin{equation}
\zeta(t,\tau) = 2 \int_\tau^t dt^\prime \varepsilon(t^\prime),
\end{equation}
which for the core-reservoir splitting considered here is defined as
\begin{equation}
\zeta(t,\tau) = 2 \int_\tau^t dt^\prime \left[ \varepsilon + \sum_{k\in {\rm res}} c_k Q_k(t) \right].
\end{equation}
The true population difference, $P(t) = \langle \sigma_z (t) \rangle$ is finally given by the average
of $\tilde{P}(t)$ over classical trajectories with initial conditions sampled from the appropriate distribution.

Importantly, the scheme described here yields the correct population dynamics of a two-level system driven by a classical external field,
so that by
taking the limit $\omega^* \rightarrow \infty$, we remove all bath modes from the core, and recover the Ehrenfest
solution.  In the opposite limit, $\omega^* \rightarrow 0$, we recover the full NIBA treatment.

Because we are not treating the core exactly, it is important to emphasize that $\omega^*$ is no longer a convergence parameter.
Instead, $\omega^*$ tunes the balance between the two approximate methods, here taken to be Ehrenfest dynamics and NIBA.  The optimal
{\em a priori} choice of this switching frequency is an interesting problem for future work, but here we present a physically
motivated prescription for its determination.  We again recall that ``adiabaticity'' is inherently a problem of timescale separation.
The adiabatic regime is physically realized when the timescale of the bath is much greater than that of the system, and {\em vice versa}
for the non-adiabatic regime.  A given bath mode $k$ could be classified into ``core'' or ``reservoir'' by comparing its
characteristic timescale, $\omega_k^{-1}$ with that of the system, $\omega_{sys}^{-1}$.  For the spin-boson Hamiltonian, Eq.~(\ref{eq:spinboson}),
the system frequency $\omega_{sys}$ is given by the Rabi frequency,
\begin{equation}
\omega_R = 2\sqrt{\varepsilon^2 + \Delta^2}.
\end{equation}
In practice we use the smooth switching function, Eq.~(\ref{eq:smoothswitch}), and following the line of argument above, set the switching
frequency equal to the Rabi frequency, i.e. $\omega^* = \omega_R$.  This procedure effectively partitions the bath modes into those
that are faster than the system dynamics (comprising the core) and those that are slower (comprising the reservoir).  While comparison
with existing exact results has shown that this choice, $\omega^* = \omega_R$, is not always truly optimal, we will show in
the next section that it nevertheless provides a very robust methodology, with quantitative predictive ability.

\section{Results}\label{sec:results}

In all our results we compare to the numerically exact population dynamics computed by Thoss, Wang, and Miller\cite{tho01}
for the spin-boson Hamiltonian, Eq.~(\ref{eq:spinboson}), with a Debye spectral density, Eq.~(\ref{eq:debye}) and initial
condition $\rho(0) = |1\rangle\langle 1| \exp(-\beta H_b)/Z_b$, i.e. $\delta = 0$ in Eqs.~(\ref{eq:kernel1})-(\ref{eq:kernel2}).

\begin{figure}[t]
\centering
\includegraphics[scale=0.4]{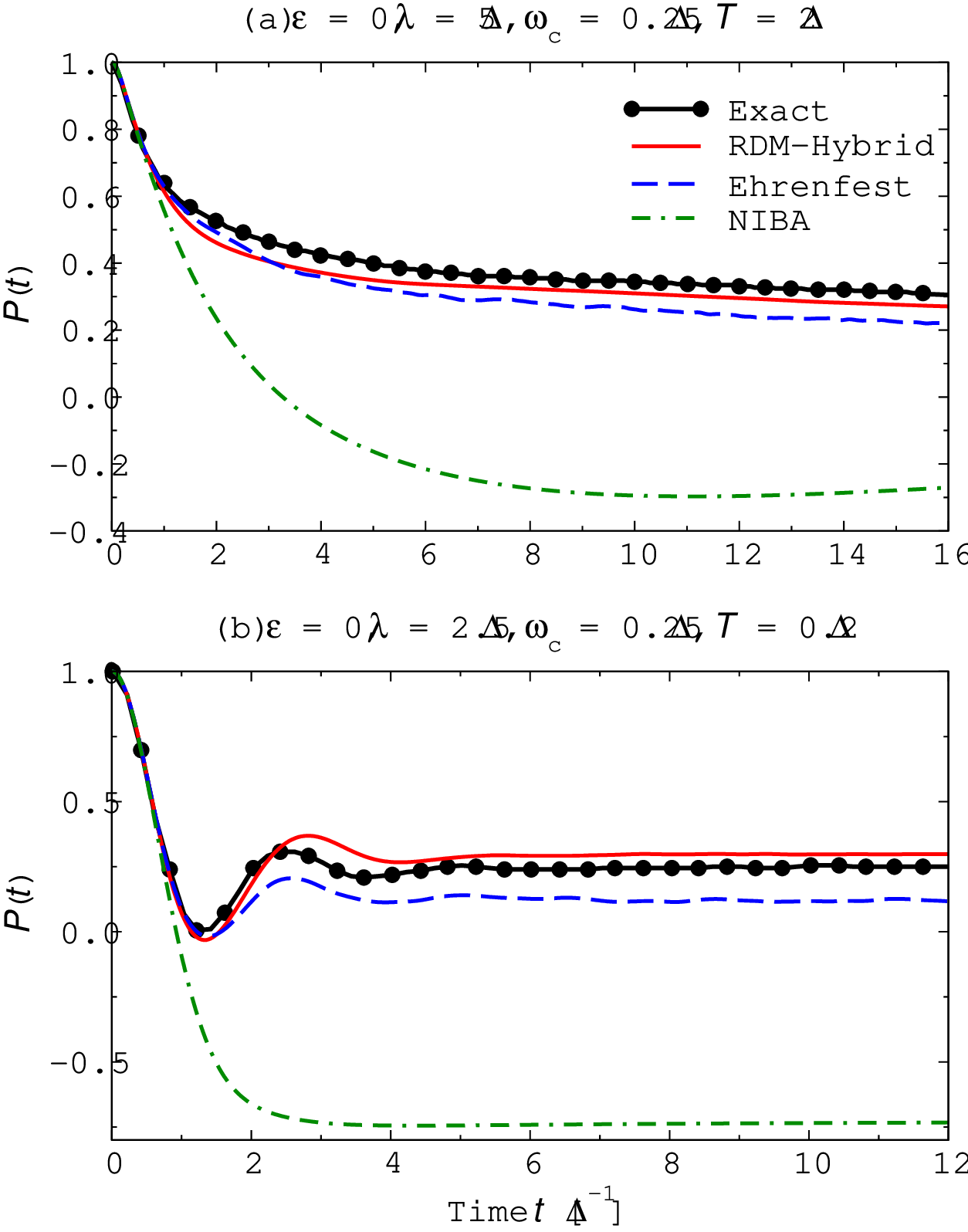}
\caption{\label{fig:adiabatic}Population dynamics of the spin-boson model in the adiabatic regime ($\omega_c/\Delta < 1$) without an energetic
bias. The bath has a Debye spectral density and parameters as given in the figure. Note the difference in the time axis.}
\end{figure}

We begin in the adiabatic regime, $\omega_c/\Delta < 1$, where Ehrenfest dynamics are known to be relatively accurate.  
Fig.~\ref{fig:adiabatic}(a) shows that this is indeed the case for high temperature, $T=2\Delta$.  Though qualitatively good, the very strong
coupling, $\lambda = 5\Delta$, degrades the accuracy of the Ehrenfest approach.  However, our RDM-Hybrid method yields a long-time population decay in
much better agreement with exact results.  In Fig.~\ref{fig:adiabatic}(b), we investigate another system with strong coupling, $\lambda = 2.5\Delta$,
but with a temperature one order of magnitude lower.  The classical mechanical approximation made by using Ehrenfest dynamics intrinsically
worsens for lower temperature, where quantum mechanical effects are expected to play a more dominant role.  We see that the RDM-Hybrid result
is again in better quantitative agreement with the exact long-time limit.
Both panels (a) and (b) in Fig.~\ref{fig:adiabatic} underscore the important point that our RDM-Hybrid approach is
not simply an ``average'' of the two methods, but can in fact generate qualitatively different dynamics which are not situated ``in between'' those of
the individual methods.

The system trapping seen above leading to a quasi-stationary state
is nearly impossible to capture with perturbative quantum master equation approaches, as exemplified by the poor NIBA results.
For comparison, the WTM MCTDH-Ehrenfest scheme required 25\% of the bath-modes to be treated
quantum mechanically, supporting our observation of largely classical bath dynamics.

\begin{figure}[t]
\centering
\includegraphics[scale=0.4]{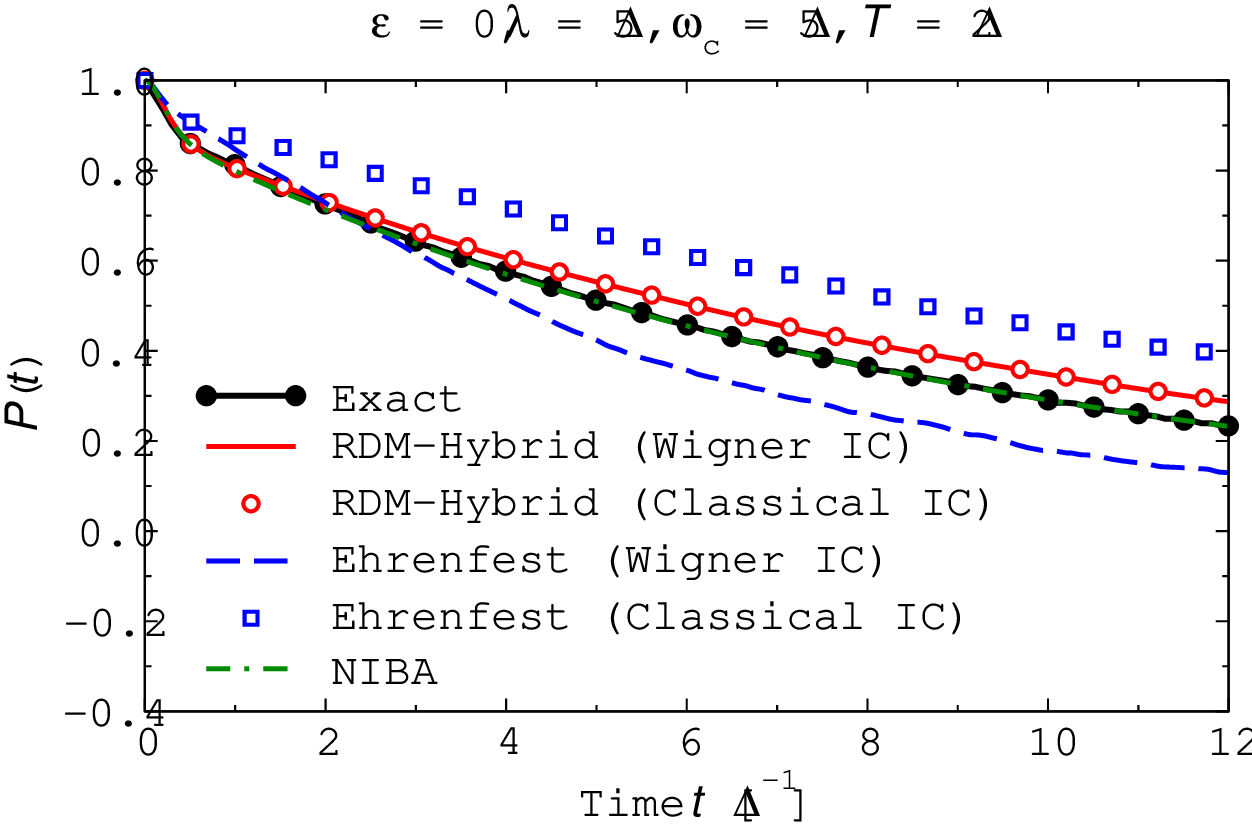}
\caption{\label{fig:nonadiabatic}Population dynamics of the spin-boson model in the non-adiabatic regime ($\omega_c/\Delta > 1$) without an energetic
bias. The bath has a Debye spectral density and parameters as given in the figure.  Also shown are the effects of classical vs. Wigner sampling
of the bath (reservoir) initial conditions for Ehrenfest dynamics (RDM-Hybrid).}
\end{figure}

In Fig.~\ref{fig:nonadiabatic}, we investigate the non-adiabatic regime, $\omega_c/\Delta > 1$, with strong coupling $\lambda = 5\Delta$,
where NIBA is quantitatively accurate.  Here we show the effects of sampling from a classical distribution or a Wigner distribution.
The Ehrenfest approximation with Wigner initial conditions yields a decay which is too rapid, while that with classical initial conditions
yields a decay which is too slow.  Both completely miss the two-step relaxation dynamics
at very short times.  RDM-Hybrid captures these short time dynamics exactly and shows a robust insensitivity to the initial sampling employed.
Although our hybrid method exhibits a slight discrepancy
at longer times, it is much more accurate than the Ehrenfest approach.

\begin{figure}[t]
\centering
\includegraphics[scale=0.4]{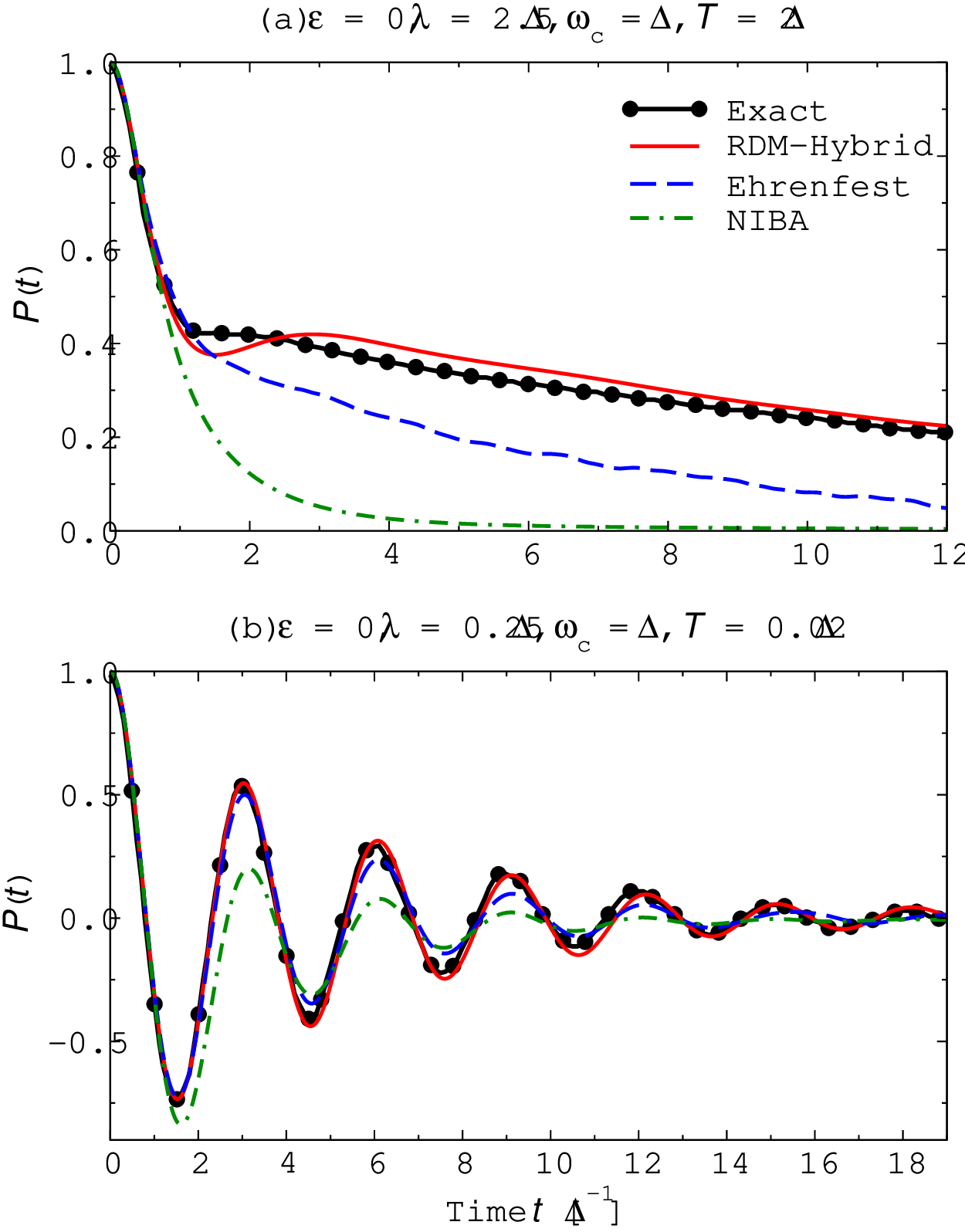}
\caption{\label{fig:intermediate}Population dynamics of the spin-boson model in the intermediate regime ($\omega_c/\Delta = 1$) without an energetic
bias. The bath has a Debye spectral density and parameters as given in the figure. Note the difference in the time axis.}
\end{figure}

Having demonstrated that our RDM-Hybrid approach can yield very good agreement with the exact results when only one of its composite
methods (Ehrenfest dynamics or NIBA) is successful, we now consider in Fig.~\ref{fig:intermediate}
the intermediate regime, $\omega_c/\Delta = 1$, where neither method is
particularly accurate by itself.  Fig.~\ref{fig:intermediate}(a) depicts the population dynamics at the high temperature $T=2\Delta$ for
relatively strong coupling, $\lambda = 2.5\Delta$.  Here, both NIBA and Ehrenfest dynamics are in very poor numerical agreement with the exact result,
though the Ehrenfest dynamics correctly predicts a two-step relaxation, albeit only qualitatively.
On the contrary, RDM-Hybrid correctly exhibits a rebound in the population dynamics near $t \sim 2\Delta^{-1}$ (although one that
is somewhat overpronounced) and yields dynamics
at all later times in excellent agreement with the exact result.

In the lower panel, Fig.~\ref{fig:intermediate}(b), we consider a weaker coupling and a temperature {\em two orders of magnitude} lower.
Here, NIBA does quite poorly and Ehrenfest dynamics are mildly better, exhibiting slightly overdamped oscillations and a weak phase-shift
at long times.  Once again, RDM-Hybrid is nearly exact, despite this severely low temperature.

In both examples above, i.e. Figs.~\ref{fig:intermediate}(a) and (b), Thoss, Wang, and Miller found it necessary to treat over
50\% of the bath modes quantum mechanically with MCTDH.  It is remarkable that our computationally inexpensive
RDM-Hybrid method employing NIBA for the core modes is able to yield such similarly accurate results for these problems,
which serves as a testament to its robustness.

\begin{figure}[t]
\centering
\includegraphics[scale=0.4]{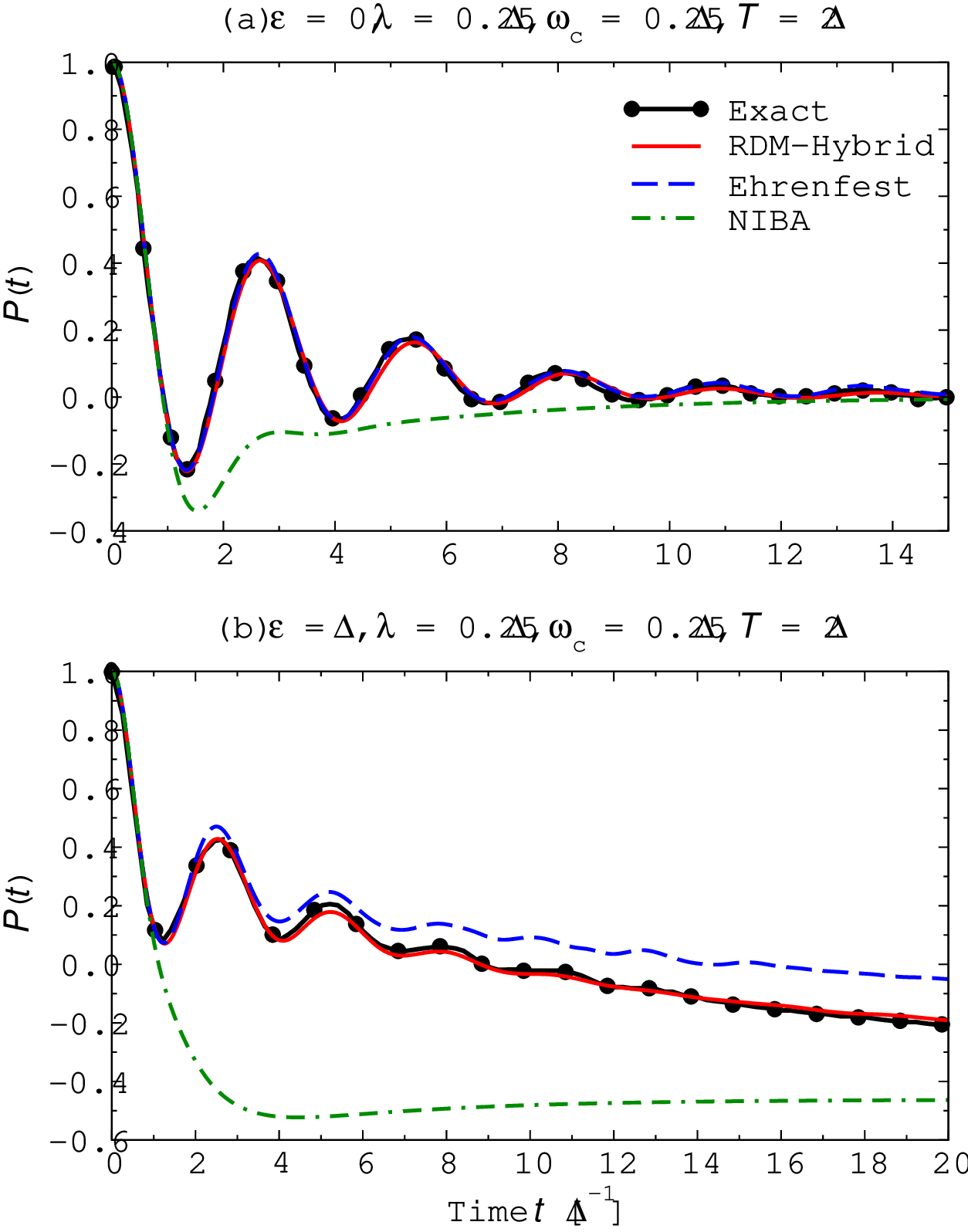}
\caption{\label{fig:adiabaticbias}Population dynamics of the spin-boson model in the adiabatic regime ($\omega_c/\Delta < 1$) without an energetic
bias (a) and with an energetic bias (b). The bath has a Debye spectral density and parameters as given in the figure. Note the difference in the time axis.}
\end{figure}

\begin{figure}[t]
\centering
\includegraphics[scale=0.4]{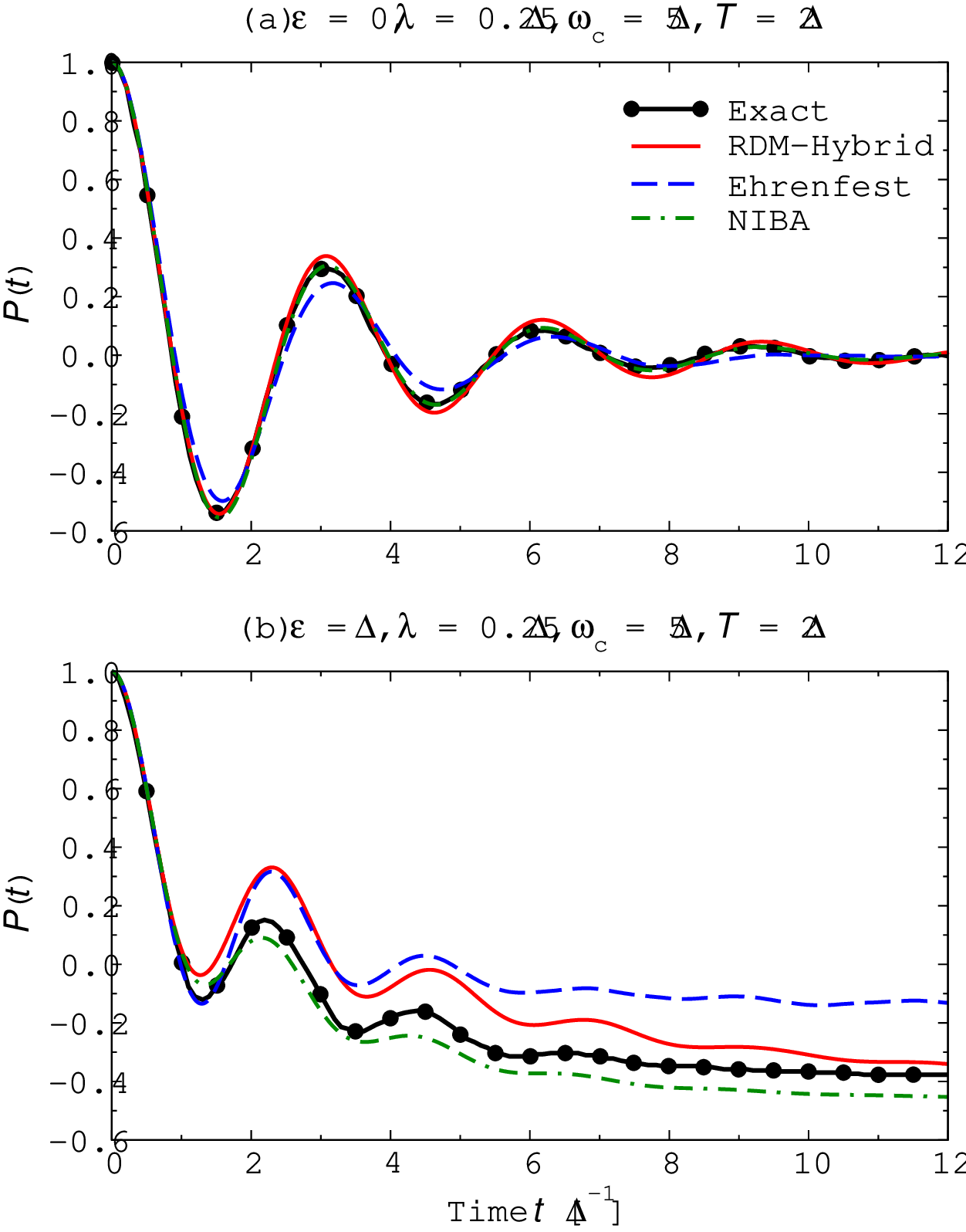}
\caption{\label{fig:nonadiabaticbias}Population dynamics of the spin-boson model in the non-adiabatic regime ($\omega_c/\Delta > 1$) without an energetic
bias (a) and with an energetic bias (b). The bath has a Debye spectral density and parameters as given in the figure.}
\end{figure}

We conclude this section by briefly considering the effect of an energetic bias on the performance of our RDM-Hybrid method.  In
Figs.~\ref{fig:adiabaticbias}(a) and (b), we show two sets of population dynamics for the same high-temperature, adiabatic
regime, but with an energetic bias $\varepsilon=\Delta$ in panel (b).  As expected, NIBA performs poorly due to the adiabaticity
of the conditions under consideration
in both panels.  As for the Ehrenfest method, while it gives excellent results in the unbiased case [panel (a)], it strongly deviates
from the correct dynamics in the presence of a bias [panel (b)].  This deficiency of the Ehrenfest approach and similar quantum-classical
methods for biased system is well-known and typically connected to unphysical treatment of zero-point energy in classical treatments.
Attempts to correct this behavior have included an adjustable zero-point energy parameter for sampling the initial conditions
\cite{mul99} and its determination by comparison with exact short-time moments.\cite{gol00_reich}

In spite of this potential concern, we see in Fig.~\ref{fig:adiabaticbias} that the RDM-Hybrid approach performs incredibly well and
equally so for both unbiased and biased systems.  Unfortunately, this success is not universal, as seen in
Fig.~\ref{fig:nonadiabaticbias}, which considers the effect of a bias on non-adiabatic dynamics.  While RDM-Hybrid is
very successful for the unbiased case, it performs more poorly in the presence of a bias.  This error appears to be attributable 
to the Ehrenfest dynamics, which RDM-Hybrid tracks quite closely at short times.  Nonetheless, the RDM-Hybrid population
does appear to approach the correct long-time value, which both Ehrenfest and NIBA incorrectly predict.

\section{Conclusions and Future Work}\label{sec:conc}
To summarize, we have formulated a quantum-classical methodology for the quantum dynamics of a coupled system-bath Hamiltonian,
averaged over the bath degrees of freedom, by employing a core-reservoir partitioning of the bath degrees of freedom.
Unlike the technique of Wang, Thoss, and Miller\cite{wan01}, our RDM-Hybrid approach
avoids the use of expensive wavefunction-based quantum mechanics, replacing it by a reduced density matrix calculation.
This allows for flexibility to use a variety of inexpensive approximate approaches for different parts of the computation,
allowing for combined accuracy and efficiency.

Within the above framework, we then used the noninteracting blip approximation (NIBA) for the reduced dynamics, yielding an efficient,
scalable quantum dynamics methodology with excellent accuracy for both adiabatic and non-adiabatic parameter regimes.
Specifically, for $N_s$ discrete states, a numerical implementation of NIBA scales only as $N_s^2$ (see Eq.~(\ref{eq:niba})),
which allows for very efficient investigation of bigger systems.
The largest
discrepancy with exact results was found in the non-adiabatic regime with an energetic bias.  To alleviate such deficiencies, we enumerate 
a number of avenues for future work.

One could imagine pursuing alternative approximate treatments of the quantum core-averaged
system RDM.  For example, the accuracy of NIBA (employed here)
is known to degrade at low temperature, in particular for biased systems.  The Redfield equations, on the other hand, have been very
successfully applied to such systems, being restricted, however, to weak-coupling and non-adiabatic baths.  Thus, an RDM-Hybrid approach employing
Redfield and Ehrenfest dynamics may be one way to extend the validity of the Redfield equations into these regimes where it would otherwise fail.

Ideally, one would like an exact treatment of the quantum system RDM, such that the method becomes numerically exact when all modes
are included in the core.  As alluded to at the end of Sec.~\ref{ssec:hybrid}, we believe the tensor propagation scheme of Makri et al.
\cite{mak92,mak94,mak95_1,mak95_2} is especially promising. Because the QUAPI algorithm scales exponentially with the
memory time required to span the bath correlation function, the partitioning discussed here leading to a reduced bath correlation time
would constitute an enormous reduction in computational expense.  Work along this direction
is currently in progress.

Lastly, for a chosen system-core dynamics method, a more rigorous investigation of the form of the switching function, $S(\omega,\omega^*)$,
and switching frequency, $\omega^*$, should be pursued.
Ideally these choices should be automated and not left up to physical considerations as was done here.
For example, one might try comparing the short-time moments of the population dynamics to those obtained exactly from quantum perturbation theory.  
The investigation of this point will be a subject of future work.

In spite of these desired improvements, the RDM-Hybrid method employing NIBA has been shown to be very accurate, often correcting
the long-time populations of the otherwise accurate Ehrenfest dynamics.
Future work will include the application and investigation of both of these methods to models
of electronic energy transfer in molecular aggregates, including the multi-site Frenkel exciton Hamiltonian describing 
the Fenna-Matthews-Olsen complex, a prototypical photosynthetic system.

\begin{acknowledgments}
The authors gratefully thank Aaron Kelly for a critical reading of the manuscript and helpful comments.
T.C.B.\ was supported by the Department of Energy Office of Science Graduate Fellowship Program (DOE SCGF),
administered by ORISE-ORAU under Contract No.~DE-AC05-06OR23100.  D.R.R.\ was supported by the National Science
Foundation under Grant No.~CHE-0719089.
\end{acknowledgments}


\begin{thebibliography}{40}%
\makeatletter
\providecommand \@ifxundefined [1]{%
 \@ifx{#1\undefined}
}%
\providecommand \@ifnum [1]{%
 \ifnum #1\expandafter \@firstoftwo
 \else \expandafter \@secondoftwo
 \fi
}%
\providecommand \@ifx [1]{%
 \ifx #1\expandafter \@firstoftwo
 \else \expandafter \@secondoftwo
 \fi
}%
\providecommand \natexlab [1]{#1}%
\providecommand \enquote  [1]{``#1''}%
\providecommand \bibnamefont  [1]{#1}%
\providecommand \bibfnamefont [1]{#1}%
\providecommand \citenamefont [1]{#1}%
\providecommand \href@noop [0]{\@secondoftwo}%
\providecommand \href [0]{\begingroup \@sanitize@url \@href}%
\providecommand \@href[1]{\@@startlink{#1}\@@href}%
\providecommand \@@href[1]{\endgroup#1\@@endlink}%
\providecommand \@sanitize@url [0]{\catcode `\\12\catcode `\$12\catcode
  `\&12\catcode `\#12\catcode `\^12\catcode `\_12\catcode `\%12\relax}%
\providecommand \@@startlink[1]{}%
\providecommand \@@endlink[0]{}%
\providecommand \url  [0]{\begingroup\@sanitize@url \@url }%
\providecommand \@url [1]{\endgroup\@href {#1}{\urlprefix }}%
\providecommand \urlprefix  [0]{URL }%
\providecommand \Eprint [0]{\href }%
\providecommand \doibase [0]{http://dx.doi.org/}%
\providecommand \selectlanguage [0]{\@gobble}%
\providecommand \bibinfo  [0]{\@secondoftwo}%
\providecommand \bibfield  [0]{\@secondoftwo}%
\providecommand \translation [1]{[#1]}%
\providecommand \BibitemOpen [0]{}%
\providecommand \bibitemStop [0]{}%
\providecommand \bibitemNoStop [0]{.\EOS\space}%
\providecommand \EOS [0]{\spacefactor3000\relax}%
\providecommand \BibitemShut  [1]{\csname bibitem#1\endcsname}%
\let\auto@bib@innerbib\@empty
\bibitem [{\citenamefont {Weiss}(2008)}]{wei08}%
  \BibitemOpen
  \bibfield  {author} {\bibinfo {author} {\bibfnamefont {U.}~\bibnamefont
  {Weiss}},\ }\href@noop {} {\emph {\bibinfo {title} {Quantum Dissipative
  Systems}}}\ (\bibinfo  {publisher} {World Scientific Publishing},\ \bibinfo
  {year} {2008})\BibitemShut {NoStop}%
\bibitem [{\citenamefont {Nitzan}(2006)}]{nit06}%
  \BibitemOpen
  \bibfield  {author} {\bibinfo {author} {\bibfnamefont {A.}~\bibnamefont
  {Nitzan}},\ }\href@noop {} {\emph {\bibinfo {title} {Chemical Dynamics in
  Condensed Phases}}}\ (\bibinfo  {publisher} {Oxford University Press},\
  \bibinfo {year} {2006})\BibitemShut {NoStop}%
\bibitem [{\citenamefont {Tully}(1998)}]{tul98}%
  \BibitemOpen
  \bibfield  {author} {\bibinfo {author} {\bibfnamefont {J.~C.}\ \bibnamefont
  {Tully}},\ }\href@noop {} {\emph {\bibinfo {title} {Classical and Quantum
  Dynamics in Condensed Phase Simulations}}},\ edited by\ \bibinfo {editor}
  {\bibfnamefont {B.~J.}\ \bibnamefont {Berne}}, \bibinfo {editor}
  {\bibfnamefont {G.}~\bibnamefont {Ciccotti}}, \ and\ \bibinfo {editor}
  {\bibfnamefont {D.~F.}\ \bibnamefont {Coker}}\ (\bibinfo  {publisher} {World
  Scientific Publishing Company},\ \bibinfo {year} {1998})\BibitemShut
  {NoStop}%
\bibitem [{\citenamefont {Stock}(1995)}]{sto95}%
  \BibitemOpen
  \bibfield  {author} {\bibinfo {author} {\bibfnamefont {G.}~\bibnamefont
  {Stock}},\ }\href@noop {} {\bibfield  {journal} {\bibinfo  {journal} {J.
  Chem. Phys.}\ }\textbf {\bibinfo {volume} {103}},\ \bibinfo {pages} {1561}
  (\bibinfo {year} {1995})}\BibitemShut {NoStop}%
\bibitem [{\citenamefont {Pechukas}(1969)}]{pec69}%
  \BibitemOpen
  \bibfield  {author} {\bibinfo {author} {\bibfnamefont {P.}~\bibnamefont
  {Pechukas}},\ }\href@noop {} {\bibfield  {journal} {\bibinfo  {journal}
  {Phys. Rev.}\ }\textbf {\bibinfo {volume} {181}},\ \bibinfo {pages} {174}
  (\bibinfo {year} {1969})}\BibitemShut {NoStop}%
\bibitem [{\citenamefont {Tully}(1990)}]{tul90}%
  \BibitemOpen
  \bibfield  {author} {\bibinfo {author} {\bibfnamefont {J.~C.}\ \bibnamefont
  {Tully}},\ }\href@noop {} {\bibfield  {journal} {\bibinfo  {journal} {J.
  Chem. Phys.}\ }\textbf {\bibinfo {volume} {93}},\ \bibinfo {pages} {1061}
  (\bibinfo {year} {1990})}\BibitemShut {NoStop}%
\bibitem [{\citenamefont {Miller}(2001)}]{mil01}%
  \BibitemOpen
  \bibfield  {author} {\bibinfo {author} {\bibfnamefont {W.~H.}\ \bibnamefont
  {Miller}},\ }\href@noop {} {\bibfield  {journal} {\bibinfo  {journal} {J.
  Phys. Chem. A}\ }\textbf {\bibinfo {volume} {105}},\ \bibinfo {pages} {2942}
  (\bibinfo {year} {2001})}\BibitemShut {NoStop}%
\bibitem [{\citenamefont {Miller}(2009)}]{mil09}%
  \BibitemOpen
  \bibfield  {author} {\bibinfo {author} {\bibfnamefont {W.~H.}\ \bibnamefont
  {Miller}},\ }\href@noop {} {\bibfield  {journal} {\bibinfo  {journal} {J.
  Phys. Chem. A}\ }\textbf {\bibinfo {volume} {113}},\ \bibinfo {pages} {1405}
  (\bibinfo {year} {2009})}\BibitemShut {NoStop}%
\bibitem [{\citenamefont {Shi}\ and\ \citenamefont {Geva}(2003)}]{shi03}%
  \BibitemOpen
  \bibfield  {author} {\bibinfo {author} {\bibfnamefont {Q.}~\bibnamefont
  {Shi}}\ and\ \bibinfo {author} {\bibfnamefont {E.}~\bibnamefont {Geva}},\
  }\href@noop {} {\bibfield  {journal} {\bibinfo  {journal} {J. Chem. Phys.}\
  }\textbf {\bibinfo {volume} {118}},\ \bibinfo {pages} {8173} (\bibinfo {year}
  {2003})}\BibitemShut {NoStop}%
\bibitem [{\citenamefont {Poulsen}, \citenamefont {Nyman},\ and\ \citenamefont
  {Rossky}(2003)}]{pou03}%
  \BibitemOpen
  \bibfield  {author} {\bibinfo {author} {\bibfnamefont {J.~A.}\ \bibnamefont
  {Poulsen}}, \bibinfo {author} {\bibfnamefont {G.}~\bibnamefont {Nyman}}, \
  and\ \bibinfo {author} {\bibfnamefont {P.~J.}\ \bibnamefont {Rossky}},\
  }\href@noop {} {\bibfield  {journal} {\bibinfo  {journal} {J. Chem. Phys.}\
  }\textbf {\bibinfo {volume} {119}},\ \bibinfo {pages} {12179} (\bibinfo
  {year} {2003})}\BibitemShut {NoStop}%
\bibitem [{\citenamefont {Bonella}\ and\ \citenamefont {Coker}(2005)}]{bon05}%
  \BibitemOpen
  \bibfield  {author} {\bibinfo {author} {\bibfnamefont {S.}~\bibnamefont
  {Bonella}}\ and\ \bibinfo {author} {\bibfnamefont {D.~F.}\ \bibnamefont
  {Coker}},\ }\href@noop {} {\bibfield  {journal} {\bibinfo  {journal} {J.
  Chem. Phys.}\ }\textbf {\bibinfo {volume} {122}},\ \bibinfo {pages} {194102}
  (\bibinfo {year} {2005})}\BibitemShut {NoStop}%
\bibitem [{\citenamefont {Dunkel}, \citenamefont {Bonella},\ and\ \citenamefont
  {Coker}(2008)}]{dun08}%
  \BibitemOpen
  \bibfield  {author} {\bibinfo {author} {\bibfnamefont {E.~R.}\ \bibnamefont
  {Dunkel}}, \bibinfo {author} {\bibfnamefont {S.}~\bibnamefont {Bonella}}, \
  and\ \bibinfo {author} {\bibfnamefont {D.~F.}\ \bibnamefont {Coker}},\
  }\href@noop {} {\bibfield  {journal} {\bibinfo  {journal} {J. Chem. Phys.}\
  }\textbf {\bibinfo {volume} {129}},\ \bibinfo {pages} {114106} (\bibinfo
  {year} {2008})}\BibitemShut {NoStop}%
\bibitem [{\citenamefont {Meyer}\ and\ \citenamefont {Miller}(1979)}]{mey79}%
  \BibitemOpen
  \bibfield  {author} {\bibinfo {author} {\bibfnamefont {H.-D.}\ \bibnamefont
  {Meyer}}\ and\ \bibinfo {author} {\bibfnamefont {W.~H.}\ \bibnamefont
  {Miller}},\ }\href@noop {} {\bibfield  {journal} {\bibinfo  {journal} {J.
  Chem. Phys.}\ }\textbf {\bibinfo {volume} {70}},\ \bibinfo {pages} {3214}
  (\bibinfo {year} {1979})}\BibitemShut {NoStop}%
\bibitem [{\citenamefont {Stock}\ and\ \citenamefont {Thoss}(1997)}]{sto97}%
  \BibitemOpen
  \bibfield  {author} {\bibinfo {author} {\bibfnamefont {G.}~\bibnamefont
  {Stock}}\ and\ \bibinfo {author} {\bibfnamefont {M.}~\bibnamefont {Thoss}},\
  }\href@noop {} {\bibfield  {journal} {\bibinfo  {journal} {Phys. Rev. Lett.}\
  }\textbf {\bibinfo {volume} {78}},\ \bibinfo {pages} {578} (\bibinfo {year}
  {1997})}\BibitemShut {NoStop}%
\bibitem [{\citenamefont {Thoss}\ and\ \citenamefont {Stock}(1999)}]{tho99}%
  \BibitemOpen
  \bibfield  {author} {\bibinfo {author} {\bibfnamefont {M.}~\bibnamefont
  {Thoss}}\ and\ \bibinfo {author} {\bibfnamefont {G.}~\bibnamefont {Stock}},\
  }\href@noop {} {\bibfield  {journal} {\bibinfo  {journal} {Phys. Rev. A}\
  }\textbf {\bibinfo {volume} {59}},\ \bibinfo {pages} {64} (\bibinfo {year}
  {1999})}\BibitemShut {NoStop}%
\bibitem [{\citenamefont {Bloch}(1957)}]{blo57}%
  \BibitemOpen
  \bibfield  {author} {\bibinfo {author} {\bibfnamefont {F.}~\bibnamefont
  {Bloch}},\ }\href@noop {} {\bibfield  {journal} {\bibinfo  {journal} {Phys.
  Rev.}\ }\textbf {\bibinfo {volume} {105}},\ \bibinfo {pages} {1206} (\bibinfo
  {year} {1957})}\BibitemShut {NoStop}%
\bibitem [{\citenamefont {Redfield}(1965)}]{red65}%
  \BibitemOpen
  \bibfield  {author} {\bibinfo {author} {\bibfnamefont {A.~G.}\ \bibnamefont
  {Redfield}},\ }\href@noop {} {\bibfield  {journal} {\bibinfo  {journal} {Adv.
  Magn. Reson.}\ }\textbf {\bibinfo {volume} {1}},\ \bibinfo {pages} {1}
  (\bibinfo {year} {1965})}\BibitemShut {NoStop}%
\bibitem [{\citenamefont {Breuer}\ and\ \citenamefont
  {Petruccione}(2002)}]{bre02}%
  \BibitemOpen
  \bibfield  {author} {\bibinfo {author} {\bibfnamefont {H.-P.}\ \bibnamefont
  {Breuer}}\ and\ \bibinfo {author} {\bibfnamefont {F.}~\bibnamefont
  {Petruccione}},\ }\href@noop {} {\emph {\bibinfo {title} {The Theory of Open
  Quantum Systems}}}\ (\bibinfo  {publisher} {Oxford University Press},\
  \bibinfo {year} {2002})\BibitemShut {NoStop}%
\bibitem [{\citenamefont {Leggett}\ \emph {et~al.}(1987)\citenamefont
  {Leggett}, \citenamefont {Chakravarty}, \citenamefont {Dorsey}, \citenamefont
  {Fisher}, \citenamefont {Garg},\ and\ \citenamefont {Zwerger}}]{leg87}%
  \BibitemOpen
  \bibfield  {author} {\bibinfo {author} {\bibfnamefont {A.~J.}\ \bibnamefont
  {Leggett}}, \bibinfo {author} {\bibfnamefont {S.}~\bibnamefont
  {Chakravarty}}, \bibinfo {author} {\bibfnamefont {A.~T.}\ \bibnamefont
  {Dorsey}}, \bibinfo {author} {\bibfnamefont {M.~P.~A.}\ \bibnamefont
  {Fisher}}, \bibinfo {author} {\bibfnamefont {A.}~\bibnamefont {Garg}}, \ and\
  \bibinfo {author} {\bibfnamefont {W.}~\bibnamefont {Zwerger}},\ }\href@noop
  {} {\bibfield  {journal} {\bibinfo  {journal} {Rev. Mod. Phys.}\ }\textbf
  {\bibinfo {volume} {59}},\ \bibinfo {pages} {1} (\bibinfo {year}
  {1987})}\BibitemShut {NoStop}%
\bibitem [{\citenamefont {F{\" o}rster}(1953)}]{for53}%
  \BibitemOpen
  \bibfield  {author} {\bibinfo {author} {\bibfnamefont {T.}~\bibnamefont {F{\"
  o}rster}},\ }\href@noop {} {\bibfield  {journal} {\bibinfo  {journal}
  {Discuss. Faraday Soc.}\ }\textbf {\bibinfo {volume} {27}},\ \bibinfo {pages}
  {7} (\bibinfo {year} {1953})}\BibitemShut {NoStop}%
\bibitem [{\citenamefont {Mak}\ and\ \citenamefont {Chandler}(1989)}]{mak89}%
  \BibitemOpen
  \bibfield  {author} {\bibinfo {author} {\bibfnamefont {C.~H.}\ \bibnamefont
  {Mak}}\ and\ \bibinfo {author} {\bibfnamefont {D.}~\bibnamefont {Chandler}},\
  }\href@noop {} {\bibfield  {journal} {\bibinfo  {journal} {Phys. Rev. A}\
  }\textbf {\bibinfo {volume} {41}},\ \bibinfo {pages} {5709} (\bibinfo {year}
  {1989})}\BibitemShut {NoStop}%
\bibitem [{\citenamefont {Mak}\ and\ \citenamefont {Chandler}(1991)}]{mak91}%
  \BibitemOpen
  \bibfield  {author} {\bibinfo {author} {\bibfnamefont {C.~H.}\ \bibnamefont
  {Mak}}\ and\ \bibinfo {author} {\bibfnamefont {D.}~\bibnamefont {Chandler}},\
  }\href@noop {} {\bibfield  {journal} {\bibinfo  {journal} {Phys. Rev. A}\
  }\textbf {\bibinfo {volume} {44}},\ \bibinfo {pages} {2352} (\bibinfo {year}
  {1991})}\BibitemShut {NoStop}%
\bibitem [{\citenamefont {Egger}\ and\ \citenamefont {Mak}(1994)}]{egg94}%
  \BibitemOpen
  \bibfield  {author} {\bibinfo {author} {\bibfnamefont {R.}~\bibnamefont
  {Egger}}\ and\ \bibinfo {author} {\bibfnamefont {C.~H.}\ \bibnamefont
  {Mak}},\ }\href@noop {} {\bibfield  {journal} {\bibinfo  {journal} {Phys.
  Rev. B}\ }\textbf {\bibinfo {volume} {50}},\ \bibinfo {pages} {15210}
  (\bibinfo {year} {1994})}\BibitemShut {NoStop}%
\bibitem [{\citenamefont {Makri}(1992)}]{mak92}%
  \BibitemOpen
  \bibfield  {author} {\bibinfo {author} {\bibfnamefont {N.}~\bibnamefont
  {Makri}},\ }\href@noop {} {\bibfield  {journal} {\bibinfo  {journal} {Chem.
  Phys. Lett.}\ }\textbf {\bibinfo {volume} {193}},\ \bibinfo {pages} {435}
  (\bibinfo {year} {1992})}\BibitemShut {NoStop}%
\bibitem [{\citenamefont {Wang}, \citenamefont {Thoss},\ and\ \citenamefont
  {Miller}(2001)}]{wan01}%
  \BibitemOpen
  \bibfield  {author} {\bibinfo {author} {\bibfnamefont {H.}~\bibnamefont
  {Wang}}, \bibinfo {author} {\bibfnamefont {M.}~\bibnamefont {Thoss}}, \ and\
  \bibinfo {author} {\bibfnamefont {W.~H.}\ \bibnamefont {Miller}},\
  }\href@noop {} {\bibfield  {journal} {\bibinfo  {journal} {J. Chem. Phys.}\
  }\textbf {\bibinfo {volume} {115}},\ \bibinfo {pages} {2979} (\bibinfo {year}
  {2001})}\BibitemShut {NoStop}%
\bibitem [{\citenamefont {Thoss}, \citenamefont {Wang},\ and\ \citenamefont
  {Miller}(2001)}]{tho01}%
  \BibitemOpen
  \bibfield  {author} {\bibinfo {author} {\bibfnamefont {M.}~\bibnamefont
  {Thoss}}, \bibinfo {author} {\bibfnamefont {H.}~\bibnamefont {Wang}}, \ and\
  \bibinfo {author} {\bibfnamefont {W.~H.}\ \bibnamefont {Miller}},\
  }\href@noop {} {\bibfield  {journal} {\bibinfo  {journal} {J. Chem. Phys.}\
  }\textbf {\bibinfo {volume} {115}},\ \bibinfo {pages} {2991} (\bibinfo {year}
  {2001})}\BibitemShut {NoStop}%
\bibitem [{\citenamefont {Meyer}, \citenamefont {Manthe},\ and\ \citenamefont
  {Cederbaum}(1990)}]{mey90}%
  \BibitemOpen
  \bibfield  {author} {\bibinfo {author} {\bibfnamefont {H.-D.}\ \bibnamefont
  {Meyer}}, \bibinfo {author} {\bibfnamefont {U.}~\bibnamefont {Manthe}}, \
  and\ \bibinfo {author} {\bibfnamefont {L.~S.}\ \bibnamefont {Cederbaum}},\
  }\href@noop {} {\bibfield  {journal} {\bibinfo  {journal} {Chem. Phys.
  Lett.}\ }\textbf {\bibinfo {volume} {165}},\ \bibinfo {pages} {73} (\bibinfo
  {year} {1990})}\BibitemShut {NoStop}%
\bibitem [{\citenamefont {Manthe}, \citenamefont {Meyer},\ and\ \citenamefont
  {Cederbaum}(1992)}]{man92}%
  \BibitemOpen
  \bibfield  {author} {\bibinfo {author} {\bibfnamefont {U.}~\bibnamefont
  {Manthe}}, \bibinfo {author} {\bibfnamefont {H.-D.}\ \bibnamefont {Meyer}}, \
  and\ \bibinfo {author} {\bibfnamefont {L.~S.}\ \bibnamefont {Cederbaum}},\
  }\href@noop {} {\bibfield  {journal} {\bibinfo  {journal} {J. Chem. Phys.}\
  }\textbf {\bibinfo {volume} {97}},\ \bibinfo {pages} {3199} (\bibinfo {year}
  {1992})}\BibitemShut {NoStop}%
\bibitem [{\citenamefont {Beck}\ \emph {et~al.}(2000)\citenamefont {Beck},
  \citenamefont {Jackle}, \citenamefont {Worth},\ and\ \citenamefont
  {Meyer}}]{bec00}%
  \BibitemOpen
  \bibfield  {author} {\bibinfo {author} {\bibfnamefont {M.~H.}\ \bibnamefont
  {Beck}}, \bibinfo {author} {\bibfnamefont {A.}~\bibnamefont {Jackle}},
  \bibinfo {author} {\bibfnamefont {G.~A.}\ \bibnamefont {Worth}}, \ and\
  \bibinfo {author} {\bibfnamefont {H.-D.}\ \bibnamefont {Meyer}},\ }\href@noop
  {} {\bibfield  {journal} {\bibinfo  {journal} {Phys. Rep.}\ }\textbf
  {\bibinfo {volume} {324}},\ \bibinfo {pages} {1} (\bibinfo {year}
  {2000})}\BibitemShut {NoStop}%
\bibitem [{\citenamefont {Kapral}(2006)}]{kap06}%
  \BibitemOpen
  \bibfield  {author} {\bibinfo {author} {\bibfnamefont {R.}~\bibnamefont
  {Kapral}},\ }\href@noop {} {\bibfield  {journal} {\bibinfo  {journal} {Annu.
  Rev. Phys. Chem.}\ }\textbf {\bibinfo {volume} {57}},\ \bibinfo {pages} {129}
  (\bibinfo {year} {2006})}\BibitemShut {NoStop}%
\bibitem [{\citenamefont {Grunwald}, \citenamefont {Kelly},\ and\ \citenamefont
  {Kapral}(2009)}]{gru09}%
  \BibitemOpen
  \bibfield  {author} {\bibinfo {author} {\bibfnamefont {R.}~\bibnamefont
  {Grunwald}}, \bibinfo {author} {\bibfnamefont {A.}~\bibnamefont {Kelly}}, \
  and\ \bibinfo {author} {\bibfnamefont {R.}~\bibnamefont {Kapral}},\
  }\href@noop {} {\emph {\bibinfo {title} {Energy Transfer Dynamics in
  Biomaterial Systems}}},\ edited by\ \bibinfo {editor} {\bibfnamefont
  {I.}~\bibnamefont {Burghardt}}, \bibinfo {editor} {\bibfnamefont
  {V.}~\bibnamefont {May}}, \bibinfo {editor} {\bibfnamefont {D.~A.}\
  \bibnamefont {Micha}}, \ and\ \bibinfo {editor} {\bibfnamefont {E.~R.}\
  \bibnamefont {Bittner}}\ (\bibinfo  {publisher} {Springer-Verlag},\ \bibinfo
  {year} {2009})\BibitemShut {NoStop}%
\bibitem [{\citenamefont {Golosov}, \citenamefont {Friesner},\ and\
  \citenamefont {Pechukas}(2000)}]{gol00_fries}%
  \BibitemOpen
  \bibfield  {author} {\bibinfo {author} {\bibfnamefont {A.~A.}\ \bibnamefont
  {Golosov}}, \bibinfo {author} {\bibfnamefont {R.~A.}\ \bibnamefont
  {Friesner}}, \ and\ \bibinfo {author} {\bibfnamefont {P.}~\bibnamefont
  {Pechukas}},\ }\href@noop {} {\bibfield  {journal} {\bibinfo  {journal} {J.
  Chem. Phys.}\ }\textbf {\bibinfo {volume} {112}},\ \bibinfo {pages} {2095}
  (\bibinfo {year} {2000})}\BibitemShut {NoStop}%
\bibitem [{\citenamefont {Golosov}, \citenamefont {Friesner},\ and\
  \citenamefont {Pechukas}(1999)}]{gol99}%
  \BibitemOpen
  \bibfield  {author} {\bibinfo {author} {\bibfnamefont {A.~A.}\ \bibnamefont
  {Golosov}}, \bibinfo {author} {\bibfnamefont {R.~A.}\ \bibnamefont
  {Friesner}}, \ and\ \bibinfo {author} {\bibfnamefont {P.}~\bibnamefont
  {Pechukas}},\ }\href@noop {} {\bibfield  {journal} {\bibinfo  {journal} {J.
  Chem. Phys.}\ }\textbf {\bibinfo {volume} {110}},\ \bibinfo {pages} {138}
  (\bibinfo {year} {1999})}\BibitemShut {NoStop}%
\bibitem [{\citenamefont {Makarov}\ and\ \citenamefont {Makri}(1994)}]{mak94}%
  \BibitemOpen
  \bibfield  {author} {\bibinfo {author} {\bibfnamefont {D.~E.}\ \bibnamefont
  {Makarov}}\ and\ \bibinfo {author} {\bibfnamefont {N.}~\bibnamefont
  {Makri}},\ }\href@noop {} {\bibfield  {journal} {\bibinfo  {journal} {Chem.
  Phys. Lett.}\ }\textbf {\bibinfo {volume} {221}},\ \bibinfo {pages} {482}
  (\bibinfo {year} {1994})}\BibitemShut {NoStop}%
\bibitem [{\citenamefont {Makri}\ and\ \citenamefont
  {Makarov}(1995{\natexlab{a}})}]{mak95_1}%
  \BibitemOpen
  \bibfield  {author} {\bibinfo {author} {\bibfnamefont {N.}~\bibnamefont
  {Makri}}\ and\ \bibinfo {author} {\bibfnamefont {D.~E.}\ \bibnamefont
  {Makarov}},\ }\href@noop {} {\bibfield  {journal} {\bibinfo  {journal} {J.
  Chem. Phys.}\ }\textbf {\bibinfo {volume} {102}},\ \bibinfo {pages} {4600}
  (\bibinfo {year} {1995}{\natexlab{a}})}\BibitemShut {NoStop}%
\bibitem [{\citenamefont {Makri}\ and\ \citenamefont
  {Makarov}(1995{\natexlab{b}})}]{mak95_2}%
  \BibitemOpen
  \bibfield  {author} {\bibinfo {author} {\bibfnamefont {N.}~\bibnamefont
  {Makri}}\ and\ \bibinfo {author} {\bibfnamefont {D.~E.}\ \bibnamefont
  {Makarov}},\ }\href@noop {} {\bibfield  {journal} {\bibinfo  {journal} {J.
  Chem. Phys.}\ }\textbf {\bibinfo {volume} {102}},\ \bibinfo {pages} {4611}
  (\bibinfo {year} {1995}{\natexlab{b}})}\BibitemShut {NoStop}%
\bibitem [{\citenamefont {Stock}\ and\ \citenamefont {M{\"
  u}ller}(1999)}]{sto99}%
  \BibitemOpen
  \bibfield  {author} {\bibinfo {author} {\bibfnamefont {G.}~\bibnamefont
  {Stock}}\ and\ \bibinfo {author} {\bibfnamefont {U.}~\bibnamefont {M{\"
  u}ller}},\ }\href@noop {} {\bibfield  {journal} {\bibinfo  {journal} {J.
  Chem. Phys.}\ }\textbf {\bibinfo {volume} {111}},\ \bibinfo {pages} {65}
  (\bibinfo {year} {1999})}\BibitemShut {NoStop}%
\bibitem [{\citenamefont {M{\" u}ller}\ and\ \citenamefont
  {Stock}(1999)}]{mul99}%
  \BibitemOpen
  \bibfield  {author} {\bibinfo {author} {\bibfnamefont {U.}~\bibnamefont {M{\"
  u}ller}}\ and\ \bibinfo {author} {\bibfnamefont {G.}~\bibnamefont {Stock}},\
  }\href@noop {} {\bibfield  {journal} {\bibinfo  {journal} {J. Chem. Phys.}\
  }\textbf {\bibinfo {volume} {111}},\ \bibinfo {pages} {77} (\bibinfo {year}
  {1999})}\BibitemShut {NoStop}%
\bibitem [{\citenamefont {Goychuk}, \citenamefont {Petrov},\ and\ \citenamefont
  {May}(1995)}]{goy95}%
  \BibitemOpen
  \bibfield  {author} {\bibinfo {author} {\bibfnamefont {I.~A.}\ \bibnamefont
  {Goychuk}}, \bibinfo {author} {\bibfnamefont {E.~G.}\ \bibnamefont {Petrov}},
  \ and\ \bibinfo {author} {\bibfnamefont {V.}~\bibnamefont {May}},\
  }\href@noop {} {\bibfield  {journal} {\bibinfo  {journal} {Phys. Rev. E}\
  }\textbf {\bibinfo {volume} {52}},\ \bibinfo {pages} {2392} (\bibinfo {year}
  {1995})}\BibitemShut {NoStop}%
\bibitem [{\citenamefont {Golosov}\ and\ \citenamefont
  {Reichman}(2000)}]{gol00_reich}%
  \BibitemOpen
  \bibfield  {author} {\bibinfo {author} {\bibfnamefont {A.~A.}\ \bibnamefont
  {Golosov}}\ and\ \bibinfo {author} {\bibfnamefont {D.~R.}\ \bibnamefont
  {Reichman}},\ }\href@noop {} {\bibfield  {journal} {\bibinfo  {journal} {J.
  Chem. Phys.}\ }\textbf {\bibinfo {volume} {114}},\ \bibinfo {pages} {1065}
  (\bibinfo {year} {2000})}\BibitemShut {NoStop}%
\end{thebibliography}
%

\end{document}